\documentclass[aps,pre,twocolumn,showpacs,showkeys,preprintnumbers]{revtex4}
\usepackage{graphicx}
\usepackage{longtable}
\usepackage{dcolumn}
\usepackage{bm}
\begin{document}
\preprint{IFT/06/03}
\title{ Universality, marginal operators, and limit cycles}
\author{ Stanis{\l}aw D. G{\l}azek }
\affiliation{ Institute of Theoretical Physics, Warsaw University, 
              ul.  Ho{\.z}a 69, 00-681 Warsaw, Poland }
\author{ Kenneth G. Wilson }
\affiliation{ Department of Physics, The Ohio State University, 
              174 West 18th Ave., Columbus, Ohio 43210-1106, USA }
\date{\today}
\begin{abstract}
The universality of renormalization group limit cycle behavior is 
illustrated with a simple discrete Hamiltonian model. A non-perturbative 
renormalization group equation for the model is soluble analytically 
at criticality and exhibits one marginal operator (made necessary by 
the limit cycle) and an infinite set of irrelevant operators. Relevant 
operators are absent. The model exhibits an infinite series of bound 
state energy eigenvalues. This infinite series approaches an exact 
geometric series as the eigenvalues approach zero - also a consequence 
of the limit cycle. Wegner's eigenvalues for irrelevant operators are
calculated generically for all choices of parameters in the model. We 
show that Wegner's eigenvalues are independent of location on the limit 
cycle, in contrast with Wegner's operators themselves, which vary 
depending on their location on the limit cycle. An example is then used 
to illustrate numerically how one can tune the initial Hamiltonian to 
eliminate the first two irrelevant operators. After tuning, the 
Hamiltonian's bound state eigenvalues converge much more quickly than 
otherwise to an exact geometric series.
\end{abstract}
\maketitle
\section{ Introduction }
\label{sec:intro}
The renormalization group has been known to be able to produce (in
principle) a limit cycle for a long time \cite{Ken}. A limit cycle is an
alternative to a fixed point, although a limit cycle necessarily implies the
existence of a fixed point for a discrete version of the renormalization
group transformation: see below. Recently, the two of us have described a
discrete, analytically soluble Hamiltonian matrix that can be analyzed using
a renormalization group procedure. The result, after an initial analysis,
was a remarkably simple form of limit cycle involving only one coupling
constant \cite{letter}. Our model can also lead to chaotic renormalization group
behavior \cite{letter}. Here, we provide a more complete analysis of the same 
model. The more complete analysis leads to renormalization group 
transformations involving either a finite set of coupling constants (with 
2, 3, 4, or more couplings, as one wishes) or in the limit to a transformation 
in a space of functions of one variable. The discrete coupling constants 
become identified with the coefficients of the function's expansion about 
the origin of its argument. The focus of the more complete analysis is 
universality and the nature of marginal and irrelevant operators (in 
Wegner's sense \cite{Wegner72}) in the model in the presence of the limit 
cycle. The possibility for chaotic behavior is disregarded here.

The limit cycle picture of the model may be helpful in qualitative
understanding of more complex systems. In particular it is already known
that a three-body Hamiltonian with short range forces exhibits a limit cycle
when there is a two-body bound state at threshold \cite{nucl}. Systems that 
might be understood approximately through the three-body limit cycle example 
include nuclear \cite{nucl}, and atomic \cite{atom} three-body systems with 
short-range interactions or many-body systems with similar interactions 
that have not been studied yet \cite{be, HB, lrs}. In the exact critical limit 
of the three-body case there is a three-body bound state spectrum that has 
an infinite sequence of states converging to zero energy as a geometric 
series; the infinite sequence of bound states was noticed \cite{Thomas, Efimov} 
well before an underlying renormalization group limit cycle was identified.

Unfortunately, physical three-body systems known to date do not have bound
states precisely at threshold, the requirement for criticality and an
infinite sequence of bound states. This makes it necessary to understand
corrections to critical behavior in the three-body case. Obtaining these
corrections is a very complex undertaking that is far from complete. One
purpose of this paper is to provide an analysis of corrections to the limit
cycle behavior due to irrelevant operators in a much simpler model to
analyze than is the three-body Hamiltonian already under investigation 
\cite{NFJG}.

A second purpose of our paper is to lay a basis for searches across quantum
condensed matter physics more broadly for evidence of limit cycle behavior.
We will stress those aspects of limit cycle behavior that are or could be
universal, and therefore present if examples of limit cycles are found in
systems other than the three-body case already under study. One very
distinctive phenomenon to be on the lookout for is the presence of geometric 
series of bound state energies (discrete energy levels) converging toward 
zero. One may have to tune a real system by applying external fields and 
stresses or introducing impurities in order to approach criticality. Therefore,
our discussion of corrections to criticality leads to an example of how one 
can tune a finite model Hamiltonian to achieve rapid approach of its
eigenvalues to their limiting behavior for low energies.

Despite the complexity of a non-perturbative limit cycle as opposed to 
a simpler fixed point, the model discussed here is largely soluble 
analytically, and even when numerical procedures are needed, they
demand relatively little in the way of computer time or complex computer
programs. To be more specific, matrix elements of the model Hamiltonian
studied here, $H_{mn} = \langle m|H|n\rangle$, are \cite{letter}
\begin{equation}
\label{H}
H_{mn}(g_N, h_N) = (E_m E_n)^{1/2}
\left(\,\delta_{mn} - g_N - i h_N s_{mn}\right) \; ,
\end{equation}
where $m$ and $n$ are integers. For $m=n$, $\delta_{mn}= 1$ 
and $s_{mn} = 0$. For $m \neq n$, $\delta_{mn}=0$ and $s_{mn} 
= (m-n)/|m-n|$. The numbers $E_n = b^n$ with $b>1$, are 
eigenvalues of the operator $H_0$, which is defined in the same 
basis to have matrix elements $\langle m|H_0|n\rangle 
= H_{mn}(0,0)$. The basis states are normalized, $\langle m| 
n\rangle = \delta_{mn}$. The energies $E_n$ of the basis states 
are limited by a cutoff $\Lambda = b^N$, so that $m, n \le N$. 
The indices can be also limited from below by a large negative
integer $M$, in order to make the Hamiltonian matrix finite. But 
the lower bound $M$ will be set equal to $-\infty$
in most of the discussion that follows.

Hamiltonians defined by Eq. (\ref{H}) have a general ultraviolet 
logarithmically divergent structure, with both real and imaginary 
parts contributing to the divergence. Therefore, the coupling 
constants $g_N$ and $h_N$ are expected to depend on the cutoff 
$N = \ln{\Lambda} / \ln{b}$ if the physical content of the 
theory, e.g., the energy spectrum, is to be independent of the 
cutoff. In fact, $g_N$ exhibits asymptotic freedom as a 
function of $\Lambda$ if $h_N = 0$. When $h_N \neq 0$, $g_N$ 
exhibits instead the limit cycle behavior (or chaos) \cite{letter}.
The universality of the limit cycle behavior is studied here. 

Section \ref{sec:RG} introduces the renormalization group (RG) 
equation in the model and sets the stage for next sections. 
Section \ref{sec:lc} describes the limit cycle and a related fixed 
point. Section \ref{sec:bs} discusses the spectrum of the Hamiltonian 
and explains how the cycle is related to a geometric sequence of 
binding energies (discrete levels) that approach zero. Then 
Section \ref{sec:mio} discusses universality and Wegner's 
marginal and leading irrelevant operators in a linearized approximation, 
with non-leading operators treated in Section \ref{sec:ho}. 
The RG analysis enables us then to show details in Section 
\ref{sec:tune} of how one can tune a Hamiltonian so that the 
limit cycle structure becomes rapidly visible in the spectrum. 
Section \ref{sec:tune} includes a numerical 
example with first two leading irrelevant operators removed 
through the tuning so that the remaining irrelevant corrections 
to the limit cycle disappear at the rate given by factor $1/512$ 
per cycle, for $b=2$. Section \ref{sec:gp} describes generic 
properties of limit cycles and Section \ref{sec:c} briefly 
concludes the paper.
 
\section{ Renormalization group }
\label{sec:RG}

The eigenvalue problem with $H$ given by Eq. (\ref{H}),
\begin{equation}
\sum_{n= -\infty}^N \, H_{mn}\, \psi_n \, = \, E \, \psi_m \; ,
\label{eve}
\end{equation}
can be solved for $\psi_m$, $m \le N$, assuming that one 
knows $E$, by using Gaussian elimination. One solves 
for $\psi_N$ in terms of all other components, $\psi_n$ 
with $n < N$. Then, one expresses $\psi_{N-1}$ in terms 
of components $\psi_n$ with $n < N-1$, and so on. After 
the first $p\,$ such steps, for negligibly small $E$, 
one is able to recognize the existence of limit 
cycles in the coupling constants $g$ and $h$ of period 
$p\,$ or less, when they occur.

The eigenstate components are re-written as $\psi_n = 
b^{-n/2} \phi_n$ for all $n \le N$, and one defines
\begin{eqnarray}
\sigma_N  = \sum_{n = -\infty}^N \phi_n \, ,  \\
\pi_{N m} = \sum_{n = -\infty}^N s_{mn} \phi_n \, , 
\end{eqnarray}
The eigenvalue condition produces then a set of equations,
\begin{eqnarray}
\left( 1 - {E \over E_m} \right) \phi_m  
-g_N \sigma_N 
-ih_N \pi_{Nm}  = 0 \, ,
\end{eqnarray}
for all integers $m \le N$. For $m=N$, $\pi_{N N} = \sigma_{N-1}$
and,
\begin{eqnarray}
\label{N}
\left( 1 -g_N - \epsilon_N \right) \phi_N  = 
 ( g_N  + ih_N ) \,\sigma_{N-1} \, ,
\end{eqnarray}
where $\epsilon_N = E/E_N$. Equation (\ref{N}) gives $\phi_N$ 
in terms of all $\phi_m$ with $m < N$ contained in $\sigma_
{N-1}$. This result is then inserted into all the remaining 
equations with $m \le N-1$, leading to the new set of equations 
with the highest energy component removed, 
\begin{eqnarray}
\left( 1 - \epsilon_m \right) \phi_m  
-g_N \sigma_{N-1} -ih_N \pi_{N-1 \, m}    \nonumber \\                      
+ \left(-g_N + ih_N \right)\,   
{1 \over 1 -g_N - \epsilon_N }\,  
\left( g_N + ih_N \right) \sigma_{N-1} = 0 \, ,  \nonumber \\
\end{eqnarray}
or
\begin{eqnarray}
\left( 1 - \epsilon_m \right) \phi_m  
-\left( g_N + {g_N^2 + h_N^2 \over 1 -g_N - \epsilon_N } \right)
\,\sigma_{N-1}  \nonumber \\
\, - \, ih_N \pi_{N-1 \, m} \, = \, 0 \, .
\end{eqnarray}
These equations appear to represent a new eigenvalue problem for a 
Hamiltonian matrix with elements,
\begin{eqnarray}
\label{H1}
H_{mn}(g_{N-1}, h_{N-1}) 
&    =   & (E_m E_n)^{1/2} \, \nonumber \\
& \times &
\left(\,\delta_{mn} - g_{N-1} - i h_{N-1} s_{mn}\right) \, ,
\nonumber \\
\end{eqnarray}
where
\begin{eqnarray}
\label{R}
 g_{N-1} & = & g_N  + \, {g_N^2 + h_N^2 
  \over 1 -g_N - \epsilon_N }    \, , \\
h_{N-1}  & = & h_N                \\
         & \equiv & h \, .
\end{eqnarray}
The new cutoff is $\Lambda_{N-1} = b^{N-1} = \Lambda/b$. 
When $N$ is reduced from $N-1$ to $N-2$, $\epsilon_N$ goes over 
to $\epsilon_{N-1}$, and so on, along with the cutoff being 
reduced by successive powers of $b$. The coupling constant 
$g_{N-p}$ obtained after $p\,$ such steps depends on $g_{N-p+1}$ 
and $\epsilon_{N-p+1}$. 

Besides the sequence of coupling constants, the RG recursion 
formula can be used to find the eigenvalues $E$. A procedure 
to do so can be used to show the connection between the universal 
RG limit-cycle behavior and the energy spectrum. In principle, 
one specifies an energy $E$ and computes the sequence of 
couplings $g_n$ for that value of $E$ from the recursion 
\begin{eqnarray}
\label{recursion}
 g_{n-1} & = & g_n  + \, {g_n^2 + h^2 
  \over 1 -g_n - E/E_n }    \, , 
\end{eqnarray}
starting from $n = N$ and iterating all the way down to $n = M+1$, 
whereby one arrives at a $g_M$ that must
satisfy the condition
\begin{eqnarray}
\label{gM}
1 \, - \, g_M \, - \, E / E_M \, = \, 0 \, ,
\end{eqnarray}
if $E$ is an eigenvalue. Although one does not know in advance 
what $E$ to pick, there exist simple search routines to find out 
values of $E$ that make the expression $1-g_M-E/E_M$ equal to zero. 
The solution to the recursion formula of Eq. (\ref{recursion})
will be developed in the next section. We will also employ 
later the reverse relation to Eq. (\ref{recursion}), 
\begin{eqnarray}
\label{reverse}
g_{n+1} & = & g_n  - \, {g_n^2  + h^2 
\over 1 + g_n - E/E_{n+1} }    \, .
\end{eqnarray}

\section{ Limit cycle and resulting fixed points }
\label{sec:lc}

Equation (\ref{recursion}) can be rewritten as, 
\begin{eqnarray}
\label{tan1}
{g_{n-1} \over h}  = 
{ (g_n /h) + h/(1-\epsilon_n) \over 1- (g_n /h) \, [h/(1-\epsilon_n)]  } \, ,
\end{eqnarray}
to exhibit the structure of a trigonometric identity,
\begin{eqnarray}
\label{tan2}
\tan{(\alpha + \beta )}  = 
{ \tan{\alpha} + \tan{\beta}  \over 1-  \tan{\alpha} \, \tan{\beta}} \, .
\end{eqnarray}
Therefore, one can introduce the angles $\alpha_n$, and $\beta_n$,
\begin{eqnarray}
\label{angle1}
\alpha_n  & = & \arctan{g_n\over h} \; , \\
\label{angle2}
\beta_n     & = & \arctan{h\over 1- \epsilon_n} \; ,
\end{eqnarray}
and obtain a simplified equation,
\begin{equation}
\label{anglerec}
\alpha_{n-1} = \alpha_n + \beta_n \; .
\end{equation}
After $p\,$ steps, one obtains
\begin{equation}
\label{an-p}
\alpha_{n-p} = \alpha_n + \gamma(E,n,n-p+1)\; ,
\end{equation}
where
\begin{equation}
\label{gamma}
\gamma(E,k,l) =  \beta_k +\,.\,.\,.\,.\,+ \beta_l\; .
\end{equation}
It is clear that the coupling constant returns to its value
after the $p\,$ steps, if 
\begin{equation}
\gamma(E,n,n-p+1) = \pi\, .
\end{equation}
Although this is unlikely for arbitrary $n$ and given $E$,
a significant simplification occurs in the recursion when 
$E \rightarrow 0$, or, equivalently, when $\epsilon_n$ is 
so small that it can be neglected. In these circumstances, 
\begin{eqnarray}
\label{bE}
\beta_n & = & \arctan{h}\, + \, {h \over 1+h^2} \, {E \over b^n} \,
+ \, { h \over (1+h^2)^2 } \, \left( {E \over b^n} \right)^2 \nonumber \\
&+& \, {h (1-h^2/3) \over (1+h^2)^3 } \, \left( {E \over b^n} \right)^3 \, 
+ \, O(E^4) \; ,
\end{eqnarray}
where $O(E^4)$ denotes terms order $E^4$ and higher. The 
simplification occurs because all angles $\beta_n$ become 
equal in the limit $E\rightarrow 0$,
\begin{eqnarray}
\beta_n     & = & \arctan{h} \\
            & \equiv & \beta \; ,
\end{eqnarray}
and 
\begin{eqnarray}
\label{Ezero}
g_{n-p\,} & = & h \tan{(\alpha_n + p \, \beta)} \nonumber \\
        & = & g_n \, ,
\end{eqnarray}
if $\beta = \pi/p\,$. Thus, one obtains a cyclic behavior of
$g_n$ with period $p\,$, if
\begin{equation}
\label{tanh}
h = \tan{(\pi/p)} \, 
\end{equation}
with an integer $p$, and if there exist eigenvalues equivalent to 
zero. The latter condition corresponds to the existence of two-body
bound state at threshold in the three-body dynamics that was 
mentioned in the introduction.
                       
Equation (\ref{tanh}) for $h$ shows how the cycle emerges for 
integer $p\, \ge 3$  \cite{letter}. $p=2$ requires infinite $h$, and 
$p = 1$ is equivalent to $p = \infty$ and $h=0$. In that case, the 
cycle is infinite and one has asymptotic freedom instead of a 
finite cycle. For rational values of $p\,$ in Eq. (\ref{tanh}), 
say, $p = p_1/p_2$ with $p_1$ and $p_2$ both integers, the sequence 
of coupling constants goes over a number of twists over the full 
period $p_2 \ge 3$. For irrational values of $p\,$ in Eq. (\ref{tanh}), 
chaotic behavior appears. We limit further discussion to integer 
$p\, \ge 3 $ in Eq. (\ref{tanh}). The sequence $g_n$, $g_{n-1}$, 
..., $g_{n-p\,+1}$, repeats itself for as long as $E$ is equivalent 
to 0. 

Equation (\ref{Ezero}) has an important implication that the 
recursion formula must simplify when one computes $g_{n-p}$ 
directly from $g_n$, because if $E=0$,
$g_{n-p}$ must equal $g_n$ regardless of the value of $g_n$. 
This result suggests that one should study the relationship 
between functions $f_n(\epsilon_n) = g_n(\epsilon_n)/h$ and 
$f_{n-p}(\epsilon_{n-p}) = g_{n-p}(\epsilon_{n-p})/h$ when 
$E \neq 0$. Note that $\epsilon_{n-p} \, = \, r \, \epsilon_n$, 
where 
\begin{equation}
\label{r}
r \, = \, b^p \, ,
\end{equation}
is the energy scale-factor associated with moving down in 
energy from $E_n$ to $E_{n-p}$, which means over the entire
cycle. But in order to derive the recursion that connects 
$f_n(x)$ with $f_{n-p}(rx)$, one needs to introduce $p\,-1$ 
intermediate functions. 

Namely, we introduce functions 
$f_{n-k}(x)$ such that $f_{n-k}(\epsilon_{n-k}) = g_{n-k}
(\epsilon_{n-k})/h$, with $\epsilon_{n-k} = b^k \epsilon_n$. 
In this notation, the recursion formula (\ref{tan1}) reads,
\begin{eqnarray}
f_{n-1}(x_{n-1})  = 
{ f_n(x_n) + z_1(x_n) \over 1 -  f_n(x_n) \, z_1(x_n)  } \, ,
\end{eqnarray}
where $x_{n-1} = b\, x_n$ and $z_1(x) = h/(1-x)$. The RG 
transformation over two steps gives 
\begin{eqnarray}
f_{n-2}(x_{n-2})  = 
{ f_n(x_n) + z_2(x_n) \over 1 -  f_n(x_n) \, z_2(x_n)  } \, ,
\end{eqnarray}
where 
\begin{eqnarray}
z_2(x)  = {z_1(x) + z_1(bx) \over 1 - z_1(x) z_1(bx)  } \, .
\end{eqnarray}
Proceeding by induction, suppose that after $k$ steps one has 
\begin{eqnarray}
\label{n-k}
f_{n-k}(x_{n-k})  = 
{ f_n(x_n) + z_k(x_n) \over 1 -  f_n(x_n) \, z_k(x_n)  } \, ,
\end{eqnarray}
and one performs one more step according to Eq. (\ref{tan1}).
The result is,
\begin{eqnarray}
f_{n-k-1}(x_{n-k-1})  = 
{ f_{n-k}(x_{n-k}) + z_1(x_{n-k}) \over 1 -  f_{n-k}(x_{n-k}) \, z_1(x_{n-k})  } \, .
\end{eqnarray}
We substitute here the assumed expression (\ref{n-k}) for 
$f_{n-k}(x_{n-k})$, and obtain,
\begin{eqnarray}
\label{n-k-1}
f_{n-k-1}(x_{n-k-1})  = 
{ f_n(x_n) + z_{k+1}(x_n) \over 1 -  f_n(x_n) \, z_{k+1}(x_n)  } \, ,
\end{eqnarray}
where 
\begin{eqnarray}
\label{zk+1}
z_{k+1}(x)  = 
{ z_k(x) + z_1(b^k x) \over 1 -  z_k(x) \, z_1(b^k x)  } \, .
\end{eqnarray}

This sequence of intermediate steps implies that after $p\,$ of them,
\begin{eqnarray}
\label{n-p}
f_{n-p\,}(x_{n-p\,})  = 
{ f_n(x_n) + z_{p\,}(x_n) \over 1 -  f_n(x_n) \, z_{p\,}(x_n)  } \, ,
\end{eqnarray}
where $z_p(x)$ is a function that can be calculated 
starting with $z_1(x) = h/(1-x)$ and using recursion 
given in Eq. (\ref{zk+1}). The point is that one must have
\begin{eqnarray}
z_{p\,}(0) \, = \, 0 
\end{eqnarray}
for the identity
\begin{eqnarray}
f_{n-p\,}(0) \, = \, f_n(0) 
\end{eqnarray}
to hold with arbitrary values of $f_n(0)$, which was observed
before as a condition for the cycle to close in the case
$E=0$ after $p\,$ steps, Eq. (\ref{Ezero}). This implies that
the Taylor series expansion for $z_{p\,}(x)$ around $x=0$, starts
with a term linear in $x$.

To give an example, we take $p\,=3$, the shortest possible
period. We focus on this example because when we discuss
tuning to criticality for matrices smaller than $100 
\times 100$, the shortest period will allow us to display 
approach to the limit cycle spectrum most clearly. We have,
\begin{eqnarray}
\label{fx3}
f_{n-3}(b^3x) &=& { f_n(x) \, + \, z_3(x)
\over  1 - f_n(x)\, z_3(x) } \, , 
\end{eqnarray}
where,
\begin{eqnarray}
z_3(x) = {a_0 + a_1 + a_2 - a_0 a_1 a_2 \over 
               1 - (a_0 a_1 + a_1 a_2 + a_2 a_0 ) } \, ,
\end{eqnarray}
and $a_j = h/(1-b^j\,x)$. For $b=2$, one obtains
\begin{eqnarray}
{z_3(x)\over h} = {(7/4) x(1-x)  \over 
       1 + x^3 - (7/4)x(1+x) } \, ,
\end{eqnarray}
which has an expansion,
\begin{eqnarray}
\label{z3}
{z_3(x)\over h} = (7/4) x + (21/16) x^2 + (343/64) x^3 + O(x^4) 
\, , \nonumber \\
\end{eqnarray}
that starts with a term order $x$, as expected. For $x=0$,
i.e., for $E=0$, one obtains a fixed point $g_{n-3k} = g_n 
= g_n^*$ for arbitrary integer values of $k$. There are 
three such fixed-modulo-cycle points in the cycle of period 
3, $g_{n-3k} = g_n^*$, $g_{n-1 -3k} = g_{n-1}^*$, and 
$g_{n-2-3k} = g_{n-2}^*$. The value of $n$ does not matter.

\section{ The bound-state spectrum for the limit cycle }
\label{sec:bs}

In this section, we show that the limit cycle leads to a spectrum 
in the form of multiple exact geometric series with eigenvalues 
in each series separated by the factor $r$ of Eq. (\ref{r}) in the 
limit $N \rightarrow \, +\infty$ and $M \rightarrow \, -\infty$. 
One of these series describes bound states and is negative. We prove 
this by showing that for fixed $E$, $M$ near enough to $ - \infty$, 
and $N$ near enough to $+\infty$, the sequence of coupling constants
$g_m(E)$ approaches, for $m$ very negative but still well above $M$, 
\begin{equation}
g_m(E) = - E \, ( 1 - 1/b) / b^m \, .
\end{equation}

The above result can be obtained from Eq. (\ref{reverse}) assuming
the eigenvalue condition $g_M(E) = 1 - E/E_M$, which means 
\begin{equation}
g_M(E) = 1 - E / b^M \, .
\end{equation}
The second term is huge in comparison to $1$ or $h^2$ for finite $E$ 
and $M \rightarrow \, -\infty$. Then Eq. (\ref{reverse}) gives,
\begin{eqnarray}
g_{M+1}(E) & = & g_M(E)  - \, {g_M^2(E)  + h^2 
\over 1 + g_M(E) - E/b^{M+1} }    \, . 
\end{eqnarray}
By inspection, $g_{M+1}(E) = g_M(E)/(1+b)$ plus corrections of 
order 1 or $h^2$. If $g_{M+n-1}(E) = -g_M(E)/a_{n-1}$, then,
neglecting terms order 1 and $h^2$ again, $g_{M+n}(E) = - 
g_M(E)/(a_{n-1} + b^n)$, which implies that $a_n = 1 + b + 
... + b^n$ and for finite $m = M+n$ one obtains
\begin{equation}
\label{gM+n}
g_{M+n}(E) = - E \, (1-1/b) / b^{M+n} \, ,
\end{equation}
as promised. But the existence of the limit cycle means that in 
the limit of large positive $n$ the coupling constants return to 
the same value after $p\,$ steps.  

Let $g_N = g^*(E)$ in the limit as $N \rightarrow \infty$ through 
steps starting with a given finite $m = M+n$ with large positive 
$n$ and $M \rightarrow -\infty$, and then sending $k$ to infinity 
through steps $m+p\,$, $m+2p\,$, $m+3p\,$, ..., $m+kp\,$, ..., etc, 
thus avoiding other values on the cycle. Now suppose that $E$ is 
increased by exactly $r = b^p$. Equation (\ref{gM+n}) says that with 
$k=1$ one has $g_{m+p\,}(rE) = g_m(E)$. But by translational invariance 
in $m$ by $p\,$, which is valid in the limit cycle, this equation 
continues to hold for all larger values $k$. In the limit of large 
positive $k$, $g_{m+kp\,}(rE) = g_{m+(k-1)p\,}(E) = g^*(E)$. This means 
$rE$ is also an eigenvalue for the same $g^*$ as $E$ is. The same
argument works for all values of $m$ within one cycle, leading 
to as many series as elements in the cycle, which equals here $p\,$.

It is necessary to actually construct $p\,$ eigenvalues not separated 
by the factor $r$ using Newton's method to determine the full set of 
$p\,$ distinct geometric series. Our numerical experience \cite
{letter} is that for any combination of $b$ and $p\,$ that we have 
tried, there is precisely one negative eigenvalue sequence and $p\,-1$ 
positive eigenvalue sequences. Besides the numerical experience with 
$1 < b \lesssim 10$, one can also consider very large $b$. In this case the 
eigenvalues are $E_n (1 - g_n)$, which translates, using Eq. (\ref{angle1}),
into $E_n ( 1 - h \tan{\alpha_n})$, so that one might expect more negative 
eigenvalue series, one for every $g_n > 1$ in one cycle \cite{letter}. 
But we observe here that with $h = \tan {\beta}$ a coupling constant 
$g_n$ can be greater than 1 when $(\pi/2 - \beta) < \alpha_n < \pi/2$, 
and this can only occur for one $n$ value in a cycle. This implies that 
there is only one negative eigenvalue series for all choices of $b$, 
$g^*$, and $p$, i.e., including the large $b$ limit. The apparent 
discrepancy between this result and Eq. (\ref{gM+n}) disappears when 
one observes that Eq. (\ref{gM+n}) is valid for $|E| \gg E_{M+n}$. 

The above analysis is based on the RG transformation and attempts
to avoid specific features of the model, besides the actual form
of the transformation and the fact that we could use it in reverse. 
But our model provides also an alternative path which takes advantage 
of the model simplicity and can be used as a crosscheck. Namely, one 
can generate energy eigenvalues as follows. First of all, for exceedingly 
large $N$, and $E$ fixed, and $M$ close to $-\infty$, there is an 
eigenvalue whenever
\begin{eqnarray}
\alpha_N + \beta_N + \beta_{N-1} + \,.\,.\,.\,+ \beta_M = 
\pi /2 + m \pi \, , 
\end{eqnarray}
with $m$ arbitrary. This means that 
\begin{eqnarray}
g_N = h \tan{[\pi/2  - \gamma(E,N,-\infty)]} \, , 
\end{eqnarray}
where
\begin{eqnarray}
\gamma(E,N,M) = \arctan{h\over 1-E/E_N} \nonumber \\
+ \arctan{h \over 1-E/E_{N-1}}  +\,.\,.\,.\,+ \arctan{h\over 1-E/E_M} \, .
\nonumber \\
\end{eqnarray}
The sum becomes infinite when $M$ goes to $-\infty$ but the sum converges
because the arc tangents go to zero proportional to $E_n$ as $n$ becomes more
negative. Now we note the identity
\begin{eqnarray}
\gamma(E, N-p, -\infty) = \gamma(rE,N,-\infty) \, .
\end{eqnarray}
This means that $E$ and $r E$ can both be eigenvalues for the same $g_N$
provided that $\gamma(E,N,-\infty) = \gamma(E, N-p,-\infty)$. This is 
certainly true if $E = 0$ and $p\, \arctan{h} = \pi$. But we can go further, 
dealing with the case of large but finite $N$. Suppose $r E_0$ is an eigenvalue 
for a given $g_N$. Then we can find an eigenvlue $E$ near $E_0$ from the 
requirement that
\begin{eqnarray}
\gamma(E,N,-\infty) =  \pi + \gamma(r E_0,N,-\infty) \, .
\end{eqnarray}
This becomes, using expansion of Eq. (\ref{bE}), 
\begin{eqnarray}
\gamma(E,N,-\infty) = \gamma(E_0,N-p,-\infty) + p\, \arctan{h} \nonumber \\
+ {h \over 1 + h^2} \left(
{E_0 \over E_{N-p+1}} + {E_0 \over E_{N-p+2}} +\,.\,.\,.\,+ 
{E_0 \over E_N} \right) \nonumber \\
 + (E-E_0) S \, ,
\end{eqnarray}
where 
\begin{eqnarray}
S = \sum_{n = M}^{N-p} {h \over   E_n[ (1-E_0 / E_n)^2 + h^2]   } \, .
\end{eqnarray}
From this result, which is valid to leading order in $E$ (replaced by $E_0$) 
and in $E-E_0$, respectively, we obtain the equation for $E- E_0$,
\begin{eqnarray}
E-E_0 &=& { -h \over 1 + h^2}\left( {E_0 \over E_{N-p+1}} \right. \nonumber \\ 
&+& \left.{E_0 \over E_{N-p+2}} +\,.\,.\,.\, + {E_0 \over E_N} \right)/S \, .
\end{eqnarray}
This gives an $E-E_0$ proportional to $E_0 / E_{N-p+1}$, which is the
behavior expected from the presence of the leading irrelevant  operator.

\section{ Marginal and irrelevant operators } 
\label{sec:mio} 

In this section we present an analysis of the two-variable renormalization
group, in which $f_{n-kp}(x)$ of Section \ref{sec:lc} is represented by a 
linear approximation in $x$ for any $k$,
\begin{equation}
f_{n-kp\,}(x_{n-kp\,}) = g^*_n/h \, + \, f^{(1)}_{n-kp\,} \, x_{n-kp\,} \, ,
\end{equation}
neglecting terms of order $x_{n-kp\,}^2$ or higher. The subscript $n$
that numbers distinct points in the cycle will be omitted to simplify 
notation wherever possible. 

The RG equations for $g^*$ and $f^{(1)}$ for a jump of $p\,$ steps are,
\begin{eqnarray}
g^* = g^*   \, , \\
\label{f1}
r\,f^{(1)}_{n-p\,}  = f^{(1)}_n + (1 + g^{*2}/h^2)\, z^{(1)}_p  \, ,
\end{eqnarray}
where $z^{(1)}_p\,$ is the coefficient of $x_n$ in the expansion of 
$z_p\,(x_n)$ in powers of $x_n$. The first step in the analysis is to 
solve Eq. (\ref{f1}) for a Hamiltonian with $g_N$ given and fixed, 
which means that $f^{(1)}_N$ is zero. For $N-n$ large, and a multiple 
of $p\,$, $f^{(1)}_n$ is fixed at a value which ensures that 
$f^{(1)}_{n-p\,}$ is the same as $f^{(1)}_n$, namely
\begin{eqnarray}
f^{(1)}_n   & = &   f^{(1)*}   \\
& = & (1 + g^{*2}/h^2)\, z^{(1)}_p\, / (r - 1) \, ,
\end{eqnarray}
which is valid for $n$ of the form $N-kp\,$ with $k$ large enough.
The solution leading from the boundary condition $f^{(1)}_N = 0$ to 
the fixed point is,
\begin{eqnarray}
f^{(1)}_{N-kp\,} & = & z^{(1)}_p\,(1 + g^{*2}/h^2) \nonumber \\
& \times & (r^{-1} + r^{-2} + ... + r^{-(k-1)}) \, .
\end{eqnarray}

Our discussion of the fixed point and operators near the fixed point 
uses the variable $x$ without any reference to $n$ or $k$, and 
the function $f(x)$ is assumed to have the generic form,
\begin{equation}
\label{flinear}
f(x) = f_0 + f_1 \, x \, .
\end{equation}
The fixed point condition is, 
\begin{eqnarray}
f(rx) =  f(x) + (1 + f_0^2/h^2) \, z^{(1)}_p \, x \, .
\end{eqnarray}
Or, written in terms of a transformation $R$,
\begin{eqnarray}
f(x) &=& R[f]  \nonumber \\
     &=& f(x/r) + (1 + f_0^2/h^2)\, z^{(1)}_p\, x/r \, .
\end{eqnarray}
This has already been solved to give a one parameter family 
of fixed points,
\begin{eqnarray}
\label{fixed}
f^*(g^*,x) = f_0^*(g^*)\, + \, f_1^*(g^*)\, x \, .
\end{eqnarray}
where
\begin{eqnarray}
f_0^*(g^*) &=& g^*/h \, , \\
f_1^*(g^*) &=& f^{(1)*}(g^*) \, .
\end{eqnarray}
  
Operators are obtained by linearizing the transformation $R$ about a fixed
point. We write,
\begin{equation}
f(x) = f^*(g^*,x) + df(g^*,x) \, ,
\end{equation}
and treat $df$ as small compared to $f$. Then,
\begin{eqnarray}
R[f^* + df] & = & f(x/r) + df(x/r) + z^{(1)}_p \nonumber \\
& \times & [1 + f_0^2/h^2 + (2f_0 df_0 /h^2) \, x/r] \, .
\end{eqnarray}

Wegner's operators are extracted from an eigenvalue condition,
\begin{eqnarray}
 L(df) \, = \, w \, df(x) \, ,
\end{eqnarray}
where $w$ is an eigenvalue and $L$ is the linearized form of $R$. 
Assuming a linear function $df$,
\begin{eqnarray}
df(x) \, = \, c_0 \, + \, c_1 \, x \, ,
\end{eqnarray}
one obtains two exquations that must be satisfied simultaneously,
\begin{eqnarray}
c_0 & = & w \, c_0 \, , \\
{1\over r}\, c_1  + 2 z^{(1)}_p\, g^* \, {1\over r} \, c_0 & = & w \, c_1 \, .
\end{eqnarray}
The eigenvalue problem has two solutions, one corresponding to a marginal
operator with $w_0 = 1$, and one corresponding to an irrelevant operator with 
$w_1 = r^{-1} = 1/b^p\,$, which is less than 1. The marginal operator has 
$c_0$ arbitrary (let it be 1) but it has a $c_1$ value also, which is 
determined given $c_0$. The irrelevant operator has $c_0$ equal to zero 
and an arbitrary value for $c_1$. Critical exponents, $\lambda_l$ for
$l$-th operator, can be read from Wegner's eigenvalues \cite{Wegner72} 
using relation, 
\begin{equation}
w_l = r^{\lambda_l} \, . 
\end{equation}
Clearly, $\lambda_0$ is 0 for the marginal operator. This operator is just the 
derivative of the fixed point function with respect to $g^*$. $\lambda_1$ is
$-1$ for the irrelevant operator. This result is obtained here generically for 
all choices of parameters in the model.

The link from operators to the behavior of eigenvalues is following.
The marginal operator causes the entire eigenvalue spectrum to shift 
while preserving the fact that successive eigenvalues have a ratio of 
$r$ as long as eigenvalues are far from either the ultraviolet cutoff 
$b^N$ or the infrared cutoff $b^M$. But as one approaches either the 
ultraviolet or infrared cutoff limits, in the energy, the eigenvalue 
ratios change in ways that can only be determined numerically, except 
for the case that $g^*$ is cycled continuously through all real values 
from $g^*$ to $\infty$, then from $-\infty$ up to $g^*$ again.  This 
complete cycle can be shown to be equivalent to the renormalization 
group transformation through $p\,$ steps already discussed.

The leading irrelevant operator, if present (as it is if one solves 
the Hamiltonian of this article for finite $N$), causes the ratios of 
eigenvalues to approach  the number $r$ as the eigenvalue itself decreases 
from its maximum near $E_N$ at a rate given by $w_1 = 1/r$ for each 
reduction of the eigenvalue by $r$. This phenomenon occurs through the 
same recursion features that we used in Section \ref{sec:bs} to show the 
existence of geometric sequences of eigenvalues, except that one has to 
look for corrections due to finite $N$. If the leading irrelevant operator 
is tuned out (see Section \ref{sec:tune}), the next operator causes the 
rate to become $1/r^2$. It will be shown in the next section that Wegner's
irrelevant operators of successive orders have eigenvalues given by
successive powers of $1/r$, which suggests that the elimination of the 
next-to-leading irrelevant operator should accelerate the rate to $1/r^3$. 
Numerical example of this behavior is shown in Section \ref{sec:tune}.

\section{ Higher-order contributions }
\label{sec:ho}

In order to illustrate numerically in Section \ref{sec:tune} how the tuning 
of a finite initial Hamiltonian to criticality can be achieved, we need to 
discuss the behavior of the higher-order terms in $f(x)$, i.e., the terms 
beyond $f_0$ and $f_1$ introduced in Eq. (\ref{flinear}). Our numerical 
example in Section \ref{sec:tune} uses $N=17$ and $M=-51$, and we choose 
the smallest possible cycle period, $p\,=3$, for the largest possible 
number of cycles to fit within the bounds of these $M$ and $N$. Accordingly, 
we focus in this section on the case $p\, = 3$ and discuss the higher-order 
terms in functions $f_{n-3k}(x)$, $f_{n-3k-1}(x)$, and $f_{n-3k-2}(x)$, 
with arbitrary $n$ and $h = \tan{\pi/3} = \sqrt{3}$. 

In principle, the choice of $g_N$ is not important, but in order to have 
the limit cycle universality pattern described in terms of integers or ratios 
of small integers, it is useful to aim at $g^* = -7$ at some value of $n < N$,
see Table \ref{tab:1} and Appendix \ref{A}. 
\begin{table}[ht]
\caption{\label{tab:1}
The one parameter family of fixed points from Eq. (\ref{fixed})  
for $p\, = 3$ and $g^* = -7$.} 
\begin{ruledtabular}
\begin{tabular}{lll}
 $n$        &  $ hf_0^* $ & $ hf_1^* $     \\ 
\hline 
$n    $     &  -7        & 13            \\
$n-1  $     &  -1/2      & 13/16         \\
$n-2  $     &   5/3      & 13/9      
\end{tabular}
\end{ruledtabular}
\end{table}

For some $n$, the fixed point Eq. (\ref{fx3}) can be written in the form,
\begin{eqnarray}
\label{f8x}
f(8x)\, = \,f(x)\,+\,z_3\,[1\, +\,f(8x)\, f(x)] \, ,
\end{eqnarray}
which can be used to generate a power series solution for the function 
$f(x)$. The constant term, which we denoted $g^*/h$, is arbitrary, but 
given $g^*$, the remaining terms are determined. We can write
\begin{eqnarray}
hf(x)& = &hf(x,g^*) \nonumber \\
&=& g^* + c_1(g^*) \, x + c_2(g^*) \, x^2 + c_3(g^*) \, x^3 + ... \, , \nonumber \\
\end{eqnarray}
and the solution for the case $g^* = -7$ is,
\begin{eqnarray}
\label{f*}
hf(x,-7) & = & - 7 + 13 x - (65 / 3) x^2 + (221 / 6) x^3 \nonumber \\
         & + & O(x^4) \, . 
\end{eqnarray}
This result is extended to arbitrary values of $b$, $p\, > 3$, and $g^*$, 
in Appendix \ref{B}.

Wegner's operators are infinitesimal functions $df(x,g^*)$ which satisfy
Eq. (\ref{fx3}) when added to the fixed-point solution for $f^*(x,g^*)$. 
Namely,
\begin{eqnarray}
{ f(x,g^*) + df(x,g^*)+ z_3 \over 1 - z_3 [f(x,g^*) + df(x,g^*)]} & - & f(8x,g^*) =
 \nonumber \\
 & = & w \, df(8x,g^*)  \, ,
\end{eqnarray}
valid to first order in $df$, with $w$ being Wegner's eigenvalue. 
Each eigenoperator begins with a different power of $x$, and the 
leading power of $x$ determines $w$, namely $w_j$ is $1/8^j$ for the 
operator that begins with the power $x^j$, where $j$ can be 1, 2, 3, 
etc. Each operator then has coefficients for all higher powers of $x$, 
just as $f(x,g^*)$ does. Key details of the supporting argument are 
given in Appendix \ref{C} for arbitrary $b$, $g^*$, and integers 
$p \ge 3$.

Extracting the linear terms, the precise  equation for determining
higher-order terms in $df$ is the following, cf. Eq. (\ref{f8x}),
\begin{eqnarray}
\label{df}
w \, df(8x,g^*) & = & df(x,g^*)\, + \, z_3(x) 
[\, df(x,g^*)\, f(8x,g^*) \nonumber \\
&+& \, f(x,g^*)\, w \, df(8x,g^*)  \,] \, .
\end{eqnarray}

\section{ Tuning to a cycle }
\label{sec:tune}

We show below how to tune parameters so that a finite Hamiltonian 
can exibit universal features of a limit cycle in its spectrum.
We consider the case of $p\, =3$ and we quote numerical results
for $b=2$.
 
Our numerical studies began with alteration of the matrix element 
$H_{N\,N}$ and attempts to use this element to obtain a fixed point 
value for the leading marginal operator already in two iterations. 
The idea was to obtain the approach to scaling with the rate of 
$1/64$ instead of only $1/8$ per cycle. Then, alterations of four 
matrix elements with largest subscripts were studied, and it was 
found that by changing only these 4 matrix elements in the Hamiltonian 
of Eq. (\ref{H}), one can also make the next-to-leading irrelevant 
operator vanish, which leads to extremely fast approach to the 
geometric scaling of the eigenvalues, clearly visible within a 
small number of cycles.

We look at mappings from parameters in the initial Hamiltonian matrix
with additional couplings in the elements $H_{N\,N}$, $H_{N\,N-1}$, 
$H_{N-1\,N}$, $H_{N\,N-1}$, to the coupling $g_{N-2}$ and its 
expansion for small $E$. Denote the relevant matrix elements as,
\begin{eqnarray}
H_{N\,N} = E_N t  \, , \\
H_{N-1\,N} = \sqrt{E_N E_{N-1}} ( u + i v) \, ,\\
H_{N\,N-1} = \sqrt{E_N E_{N-1}} ( u - i v) \, ,\\
H_{N-1,N-1} = E_{N-1} z \, .
\end{eqnarray}
Then, $g_{N-2}$ can be written as the ratio of two determinants. The 
numerator is the determinant of a $3 \times 3$ matrix that can be 
written with overall energy factors $E_N$, etc., removed. Then, the 
$3 \times 3$ matrix is,
\begin{equation}
\left[
\begin{array}{ccc}
t - x /b^2  &  u-iv       &     -g-ih  \\
u+iv        &  z - x / b  &     -g-ih  \\
-g+ih       & -g+ih       &     -g
\end{array}
\right]
\end{equation}
where  $g$ is $g_N$ and $x$ is $E / E_{N-2}$.
The denominator is the determinant of the upper left $2 \times 2$ 
matrix within this $3 \times 3$ matrix. The result is that
\begin{eqnarray}
g_{N-2} &=& g - (g^2 + h^2) \nonumber \\
& \times & {  2 u - (t + z) + x (b + 1) / b^2  \over
  tz - (u^2 + v^2) - x (t b + z) / b^2 + x^2 / b^3 }  \, .\nonumber \\
\end{eqnarray}
It is convenient to set $t=u=z$, leaving $z$ and $v$ as free parameters. Then,
with $h = \sqrt{3}$,
\begin{eqnarray}
g_{N-2} & = & g \, + \, x \, (g^2 + 3)\, { b + 1 \over b^2 v^2 } \nonumber \\
& - &
x^2\, (g^2 + 3)\, z\, \left( {b + 1 \over b^2 v^2 } \right)^2\, + O(x^3) \, .
\end{eqnarray}
The fixed point value for $g_{N-2}$, through order $x^2$, can also be 
worked out, and gives
\begin{eqnarray}
g_{N-2} & = & g \, + \, x \, (g^2 + 3) { 1 \over 4(b - 1) }\nonumber \\
& + &  x^2 \, (g^2 + 3)\, 
{g + (b - 1) / (b + 1) \over 16(b - 1)^2 } \, + O(x^3) \, . \nonumber \\
\end{eqnarray}
From matching these two formulae, one obtains
\begin{eqnarray}
v^2 = 4 (b^2 - 1) / b^2 = 3 \, ,
\end{eqnarray}
and,
\begin{eqnarray}
z = - g - (b - 1) / (b + 1) = 20 / 3 \, ,
\end{eqnarray}
with the numerical values valid for $g = -7$ and $b = 2$.
\begin{table}[ht]
\caption{\label{tab:2} The left column contains eigenvalues
of the initial Hamiltonian which is tuned so that the leading 
and next-to-leading irrelevant operators reach 0 already at 
$N-2$ (see the text for details, $t=u=z= 20/3$, $v = \sqrt{3}$, 
$g_N = -7$, $p\, = 3$, $b=2$, $N=17$, $M \le -51$). The 
eigenvalues are ordered by magnitude, and the negative ones are 
separated from the positive. The middle column contains the ratio 
to prior (smaller in magnitude) eigenvalue. The right column
indicates the factor $w_3^{-1} \sim r^3 $ with which corrections 
to the ratio $r = 8$ disappear when binding energies approach 0.
The number $\sim 400$ in the right column turns into 511.6 
when $M= -51$ is changed to $M= -60$.}
\begin{ruledtabular}
\begin{tabular}{rrr}
 $E$ & ratio & $w_3^{-1}$ \\ 
\hline 
             0.115359460521   &                     &   \\
             0.326274994387   &     2.828333219599  &   \\
             0.922875684170   &     2.828521033011  &   \\
             2.610199955097   &     2.828333219599  &   \\
             7.383005473358   &     2.828521033011  &   \\
            20.881599640790   &     2.828333219600  &   \\
            59.064043787554   &     2.828521033043  &   \\
           167.052797170395   &     2.828333220314  &   \\
           472.512353132001   &     2.828521049247  &   \\
          1336.422560688180   &     2.828333591344  &   \\
          3780.110931254790   &     2.828529719903  &   \\
         10692.223889183900   &     2.828547649430  &   \\
         30306.851247105530   &     2.834475929536  &   \\
         91327.456629399810   &     3.013426102394  &   \\
       1824274.370928830000   &    19.975092247797  &   \\ 
\hline
            -3.321031582839   &                     &   \\
           -26.568252662770   &     8.000000000017  & $\sim 400$ \\
          -212.546021531960   &     8.000000008649  &  511 \\
         -1700.369094987040   &     8.000004341325  &  502 \\
        -13606.149375382800   &     8.001879953885  &  433 \\
       -111890.939577163000   &     8.223556605927  &  119 \\
\end{tabular}
\end{ruledtabular}
\end{table}

Using $t=u=z= 20/3$, $v = \sqrt{3}$, and $g_N = -7$, we find that 
\begin{eqnarray}
g_{N-2}(x) & = & -7 + 13x/(1+5x/3 -x^2/24) \nonumber \\
& = & -7\, +\, 13\, x\, - \,{65\over 3} \,x^2 \,+ o(x^3) \, ,
\end{eqnarray}
in agreement with Eq. (\ref{f*}), and indeed the eigenvalues 
converge very fast towards a fixed ratio, as illustrated by 
Table \ref{tab:2}. There is a negative eigenvalue sequence 
separated by a factor rapidly approaching 8 and a positive 
eigenvalue sequence whose ratios rapidly approach an alternating 
sequence of two ratios both near the square root of 8. As a 
numerical check, we have verified that the sum of these eigenvalues 
is to twelve decimal places the trace of the original Hamiltonian. 
The eigenvalues in Table \ref{tab:2} are given for $N = 17$ and 
$M = -51$ (low enough to have no impact on these numbers: smaller 
eigenvalues are omitted). The number $\sim 400$ in the right column 
turns into $511.6$ when $M= -51$ is changed to $M= -60$. This 
result merely confirms that the numbers in the table correspond 
to the limit cycle with $M \rightarrow -\infty$. The ratios for 
the positive eigenvalues can be compared with $b^{p\,/(p\,-1)} = 
\sqrt{8} \sim 2.82842712474619 $, looking at the average value 
$(2.828333219599 + 2.828521033011)/2 = 2.828427126305$, and at the product 
$ 2.828333219599 \, \times \, 2.828521033011 \sim 7.999999999999491$.

\section[foaltfix]{ Generic Properties of Limit Cycles }
\label{sec:gp}

The model Hamiltonian discussed in this paper has an extraordinarily simple
solution despite its exhibiting a limit cycle. The initial coupling $g_N$ is
given as an explicit analytic expression as a function of an energy
eigenvalue $E$ that involves nothing worse than a convergent infinite sum of
arc tangents. The renormalization group transformation for the model can be
written in terms of one, two, or an arbitrary number of coupling parameters.
Analysis of the transformation with $j$ couplings yields exact results for the
limit cycle itself and for the critical exponents of the $j$ leading
operators, including a marginal operator and a succession of irrelevant
operators with integer exponents.

Obviously, this is too simple to all carry over to more general renormalization 
group limit cycles. What does carry over to the more general case? First of all, 
we cannot guarantee that a renormalization group limit cycle will lead to any 
bound states. But if there are bound states close to threshold, then these bound 
states must form an infinite geometric series converging on zero. The reason for 
this infinite geometric series is the existence of a fixed point Hamiltonian for 
a change of scale $r$ that corresponds to a transformation around the complete 
limit cycle ($r = b^p$ in the discrete model). The proof of this statement begins 
with the fact that if the fixed point Hamiltonian with cutoff $\Lambda$ has an 
energy eigenvalue $E$ much less than $\Lambda$, then the Hamiltonian after the 
renormalization group transformation has cutoff $\Lambda / r$, yet exactly the 
same form. Therefore the Hamiltonian after the transformation must have an eigenvalue 
$E/r$. But the definition of the transformation is that it preserves eigenvalues, 
which means that the Hamiltonian before the transformation must have $E/r$ as an
eigenvalue too, as well as its original eigenvalue $E$. Then by following the
same reasoning over again, one can establish that the complete geometric
series $E$, $E/r$, $E/r^2$, etc., must all be eigenvalues for the fixed point
Hamiltonian.

The second observation to make is that marginal and irrelevant operators
vary at different locations on the limit cycle, but the critical exponents
of these operators do not vary. In the model, the operators have the form of
a set of coefficients $c_1(g^*_n)$ , $c_2(g^*_n)$, etc., that depend on the values
$g^*_n$ that make up the limit cycle. The critical exponents are the integers
$-1$, $-2$, etc., and do not depend on the location $n$ on the cycle. Let us 
us then consider the general case. Suppose there is a non-linear transformation 
$R_b$ that forms the basic (this is why we introduce the subscript $b$) 
renormalization group transformation that when iterated $p\,$ times, $R_b\,R_b
\,.\,.\,.\,R_b$, produces the transformation $R$ defined earlier that corresponds 
to a complete circuit of the cycle, which we could denote here by $R_r$. Suppose 
then that one has a limit cycle of Hamiltonians $H^*_n$ of period $r=b^p\,$. 
Suppose that $dH_n$ is one of Wegner's operators for this limit cycle. We now
apply the renormalization group transformation $R_r$ to $H^*_n + dH_n$ $k$ times, 
where $k$ is arbitrarily large. Suppose that the critical exponent for $dH_n$ 
were an exponent $\lambda_n$ dependent on $n$, i.e., we would have different
exponents for different $n$s. Then there would exist two ways to apply the iterated 
transformation that would give two different results, while on the other hand
they would have to give the same result in the case of limit cycle. The
first way is to replace $R_b$ iterated $kp\,$ times by $R_r$ iterated $k$
times, in which case $H^*_n + dH_n$ becomes $H^*_n + r^{\lambda_n} dH_n$. 
The second way is to apply $R_b$ to $H^*_n + dH_n$, then apply the transformation 
$R$ for $k-1$ times, then apply $R_b$ for $p\,-1$ times. The only source of $k$ 
dependence is the $k-1$ applications of $R_r$, but it is applied to $H^*_{n-1} 
+ dH_{n-1}$ which means that $dH$ is multiplied by $r^{(k-1)\lambda_{n-1}}$. 
If $\lambda_{n-1}$ is different from $\lambda_n$, there is an immediate 
contradiction in the $k$-dependence of the two different approaches. The same
argument works for all values of $n$ within the cycle.

The third observation is that the model has a marginal operator, that is,
an operator with a critical exponent of zero. The presence of a marginal
operator is guaranteed if one has a renormalization group in differential
form so that the limit cycle must be continuous, and position on the cycle
is labeled by a continuous parameter, just as $g^*$ is continuous in the model.
In this case, the marginal operator is obtained by differentiating the limit
cycle Hamiltonians themselves with respect to the parameter , and it is
easily established that the critical exponent for this operator must be
zero. But when the renormalization group is discrete, and the limit cycle is
a finite set of points rather than a continuous curve in the space of
Hamiltonians, we do not know whether or not a marginal operator has to
occur. In addition, while our model has no relevant operators, this need not
be true for other limit cycles, and indeed, the limit cycle for the
three-body case has a relevant operator associated with infinitesimal
changes in the strength of the two-body potential in the model. When the
relevant operator is absent, there is a two-body bound state exactly at
threshold, and inclusion of the relevant operator causes this threshold
bound state either to acquire a non-zero binding energy or to blend into the
two-body scattering continuum.

Finally, we showed that parameters in an initial Hamiltonian could be tuned
to eliminate the two leading irrelevant operators and thereby ensure a very
rapid approach to a pure geometric series of the bound state eigenvalues for
the Hamiltonian. We cannot guarantee that such tuning will be possible in
other examples of renormalization group limit cycles, such as the limit
cycles for the three-body Hamiltonians \cite{nucl, atom, be, Thomas, Efimov}. 
The problem is that there is a non-linear mapping from parameters in an 
initial Hamiltonian to coefficients of Wegner's operators, and with non-linear 
mappings, there is no assurance that any given set of output parameters can 
be obtained from the mapping. We were lucky to find an easily identified 
solution for the tuning problem for the starting Hamiltonian that we used.
In other cases, one has to make careful studies to establish if a desired
option for tuning can work. 

\section{ Conclusion }
\label{sec:c}

The renormalization group limit cycle discussed in this paper is
extraordinarily simple to analyze thanks to the analytic solution to the
recursion formula for $g_n$ and to the analytic formula expressing $g_N$ in
terms of any eigenvalue $E$. It is far simpler to study than the three-body
Hamiltonian of references \cite{nucl, atom, be, Thomas, Efimov}, which more 
than compensates for its lack of immediate practical applications. We do not 
know how many other examples of Hamiltonians with renormalization group limit 
cycles will be found in the future. But hopefully, the clarification of the 
expected behavior of renormalization group limit cycle behavior described in 
this article will make it easier to search for and make sense of limit cycle 
behavior in more complex circumstances. Moreover, the possibility of bound 
states forming geometric series converging on zero energy provides a startling 
characteristic to search for. 

\appendix

\section{ Calculation of entries for Table \ref{tab:1} 
for integers $p\, \ge 3$}
\label{A}

Suppose that, 
\begin{equation}
g_n = g_n^* + c_n \epsilon_n \, ,
\end{equation}
where $g_n^* = hf_0^*$, and $c_n = hf_1$. We have,
\begin{eqnarray}
g_{n-1}* + c_{n-1} \epsilon_{n-1}
& = & g_{n}^* + c_n \epsilon_n   \nonumber \\
& + & \, {( g_n^* + c_n \epsilon_n)^2 + h^2 \over 
1 - (g_n^* + c_n \epsilon_n) - \epsilon_n  } \, .
\end{eqnarray}
Expanding the right-hand side in powers of $\epsilon_n$,
and taking into account the cycle relation (\ref{tan1}) 
between the coupling constants $g_{n-1}^*$ and $g_n^*$
for $E=0$, the result is,
\begin{eqnarray}
\label{n1}
c_{n-1} \, = \,x_n \, + \,y_n \,c_n\, ,
\end{eqnarray}
with 
\begin{eqnarray}
\label{m1}
x_n & = &  
{ 1 \over  b}\,{ g_n^{*2} + h^2 \over (1-g_n^*)^2} \, , \\
y_n & = &
{ 1 \over b}\,{1 + h^2 \over (1-g_n^*)^2} \, .
\end{eqnarray}
$g_n^*$ runs over a cycle and $h$ and $b$ are constants.
Therefore, $c_n$ also approaches a certain cycle when 
$n$ is lowered. This can be seen after $p\, $ steps,
\begin{eqnarray}
c_{n-p} \, = \, u_n\, + \, v_n \, c_n \, ,
\end{eqnarray}
with 
\begin{eqnarray}
\label{un}
u_n 
& = & x_{n-p\,+1} \,\, + \,\, y_{n-p\,+1}\, x_{n-p\, +2} \nonumber \\
& + &  
y_{n-p\,+1}\, y_{n-p\,+2}\, x_{n-p\,+3} \,\, + \,\,\,.\,.\,.\,\, \nonumber \\
& + & 
y_{n-p\,+1}\, y_{n-p\,+2}\,\,.\,.\,.\,\,y_{n-1}\, x_n \, , \\
\label{vn}
v_n & = & \,  y_{n-p\,+1}\, y_{n-p\,+2}\,\,.\,.\,.\,\, y_{n-1}\,y_n \, .
\end{eqnarray}
The fixed point of this transformation over the cycle is given by
\begin{eqnarray}
\label{fixedmu}
c_n^* \, = \, u_n /(1 - v_n) \, ,
\end{eqnarray}
and the $p\,-1$ values in between can be calculated for given 
$g_n^*$ at one value of $n$ using Eq. (\ref{n1}).

\section{ Direct energy expansion in the model }
\label{B}

Including corrections due to $E \neq 0$ in Eq. (\ref{bE}), and
inserting them in Eq. (\ref{an-p}), one obtains for 
$p\, \arctan{h} = \pi$ that,
\begin{eqnarray}
\label{ga}
{g_{n-p}\over h}&=&\tan{[\alpha_n + e_1 E + e_2 E^2 + e_3 E^3 + O(E^4)]}, 
\nonumber \\
\end{eqnarray}
where $g_n$ and $\alpha_n$ are functions of $E$, and 
\begin{eqnarray}
e_1 & = & {h \over 1+h^2} \, { b_1 \over b^n} \, , \\
e_2 & = & {h \over (1+h^2)^2 } \, { b_2 \over b^{2n} }  \, , \\
e_3 & = & {h \over (1+h^2)^3 }\left(1-{h^2\over 3}\right) \, { b_3 \over b^{3n} }  \, ,
\end{eqnarray}
with $b_k = (b^{kp} - 1)/(b^k - 1)$.
Then, the coupling constant dependence on $E$ obeys the recursion,
\begin{eqnarray}
\label{gaE}
g_{n-p} & = & g_n + \tilde e_1 E + \tilde e_2 E^2 + \tilde e_3 E^3 + O(E^4) \, , \\
\tilde e_1 & = & {g_n^2 + h^2 \over 1+h^2}\, {b_1\over b^n} \, , \\
\tilde e_2 & = & {g_n^2 + h^2 \over (1+h^2)^2}\,\left[b_2 \,
+ \,g_n \, b_1^2 \right]\, {1\over b^{2n}} \, , 
\nonumber \\
          \\
\tilde e_3 & = & {g_n^2 + h^2 \over (1+h^2)^3 }
\,\left[ \left(1-{h^2\over 3}\right) \, b_3 
+ 2 g_n b_2 b_1 \right. \nonumber \\ 
& + & \left. (g_n^2 + h^2/3) b_1^3 \right]\, {1 \over b^{3n} }  \, .
\end{eqnarray}
Assuming that $f_n(rx) = g_{n-p}(rE)$ and $g_n(E) = f_n(x)$ with $x = E/E_n$,
and expanding Eq. (\ref{gaE}) in a series of powers of $x$, one arrives at 
(the subscript $n$ is dropped, but there are $p\,$ distinct functions with 
$p\,$ different values of $g^*$, forming a cycle),
\begin{eqnarray}
f(x) & = & g^* \, + \, c_1^*\, x \, + \, c_2^* \, x^2 \, + \, c_3^* \, x^3 + O(x^4) \, ,
\end{eqnarray}
with the coefficients,
\begin{eqnarray}
c_1^* & = &  {g^{*2} + h^2 \over 1+h^2}\,{b_1 \over r-1}\, ,  \\
c_2^* & = &  {g^{*2} + h^2 \over (1+h^2)^2}\,
\left[ g^* \left({b_1 \over r-1}\right)^2 + {b_2 \over r^2 - 1} \right] \, ,  \\ 
& & \nonumber \\
c_3^* & = &  {g^{*2} + h^2 \over (1+h^2)^3}\,
\left[ (g^{*2} + h^2/3)\left({b_1 \over r-1}\right)^3 \right. \nonumber \\
& & \nonumber \\
&+& \left. 
2 g^* \, {b_1 \over r-1} \, {b_2 \over r^2-1} \,  
+ \, { (1-h^2/3)b_3 \over r^3-1} \right]
\end{eqnarray}
These expressions provide fixed-modulo-cycle values of 3 additional 
coupling constants for arbitrary choices of $g^*$, $h = \tan{(\pi/p\,)}$, 
and $b$, with integer $p\, > 3$, in addition to the results obtained 
in Section \ref{sec:mio}. The $p\,$ different constants $g^*$ in the 
limit cycle, $g_n^*$, $g_{n-1}^*$, . . ., $g_{n-p+1}^*$, are determined 
by a choice of any one of them with an arbitrarily selected $n$.  
\section{ Eigenvalues $w_j$ for $j > 1$ }
\label{C}
Equation (\ref{an-p}) implies
\begin{eqnarray}
\alpha_{n-p}(rx) = \alpha_n(x) + \gamma(x) \, ,
\end{eqnarray}
where 
\begin{eqnarray}
\gamma(x) = \sum_{k=0}^{p-1} \, \arctan{h\over 1- b^k x} \, ,
\end{eqnarray}
see Eq. (\ref{gamma}). One has $\gamma(0)= \pi$, and the fixed point 
relation can be written as, 
\begin{eqnarray}
\alpha^*(rx,g^*) = \alpha^*(x,g^*) + \gamma(x) - \gamma(0) \, .
\end{eqnarray}
The coefficients of the Taylor series for the function 
$\alpha (x,g^*)$ around zero,
\begin{eqnarray}
\alpha (x,g^*) = \tau_0(g^*) + \sum_{m=1}^\infty \tau_m(g^*) x^m /m! \, ,
\end{eqnarray}
have fixed point values given by $\tau_m(g^*) = \tau^*_m(g^*)$, with 
\begin{eqnarray}
\tau^*_m(g^*) = \gamma_m /( r^m -1 ) \, ,
\end{eqnarray}
where $\gamma_m$ are given by the expansion of the known
function $\gamma(x)$,
\begin{eqnarray}
\gamma(x) = \pi + \sum_{m=1}^\infty \gamma_m x^m /m! \, .
\end{eqnarray}
Suppose that 
\begin{eqnarray}
\alpha (x,g^*) = \alpha^*(x,g^*) + \delta_j x^j \, ,
\end{eqnarray}
with infinitesimal $\delta_j$ and $j >1$. The RG transformation 
over the cycle gives,
\begin{eqnarray}
w_j \delta_j = r^{-j} \delta_j \, ,
\end{eqnarray}
and a whole series $df_j(x,g^*)$ of powers of $x$ greater than 
or equal to $j$ is generated, by Wegner's eigenvalue condition, 
\begin{eqnarray}
\label{ev}
w_j \, df_j(rx,g^*) & = & df_j(x,g^*)\, + \, z_p(x) 
[\, df_j(x,g^*)\, f(rx,g^*) \nonumber \\
&+& \, f(x,g^*)\, w_j \, df_j(rx,g^*)  \,] \, ,
\end{eqnarray}
with $f(x,g^*) = h \tan{[\alpha^*(x,g^*)]}$. 
 
\end{document}